\documentclass[11pt]{article}

\usepackage[T1]{fontenc}
\usepackage[utf8]{inputenc}
\usepackage{lmodern}
\usepackage[a4paper,margin=2.3cm]{geometry}
\usepackage{graphicx}
\usepackage{microtype}
\usepackage[round,authoryear]{natbib}
\usepackage{enumitem}
\usepackage{caption}
\usepackage{placeins}
\usepackage{xurl}
\usepackage[hidelinks]{hyperref}

\hypersetup{
  pdftitle={Why does AI unlock new possibilities in STEM education? A Bibliometric Analysis of Trends and Future Agenda},
  pdfauthor={Jesse Yusuf Chan (Zexi Chen), Mengyao Chen, Yang Hong, Ziyun Song, Haoming Wang, Xianlong Xu}
}

\setlist{nosep}
\newcommand{\email}[1]{\href{mailto:#1}{\nolinkurl{#1}}}

\title{Why does AI unlock new possibilities in STEM education?\\
\large A Bibliometric Analysis of Trends and Future Agenda}

\author{
Jesse Yusuf Chan (Zexi Chen)$^{1}$, Mengyao Chen$^{1}$, Yang Hong$^{1}$,\\
Ziyun Song$^{2}$, Haoming Wang$^{1,*}$, and Xianlong Xu$^{1,*}$\\[0.55em]
\small $^{1}$East China Normal University\\
\small $^{2}$Nantong University\\[0.4em]
\footnotesize \email{51274108057@stu.ecnu.edu.cn}\quad
\email{51274108033@stu.ecnu.edu.cn}\\
\footnotesize \email{HongYang387@outlook.com}\quad
\email{drivingintosea@gmail.com}\\
\footnotesize \email{wfrank0222@gmail.com}\quad
\email{xlxu@eec.ecnu.edu.cn}
}

\date{}

\begin{document}
\maketitle

\begin{abstract}
STEM education faces challenges in personalization and interdisciplinary integration. AI technology has brought new possibilities, but the mechanisms by which AI reshapes the STEM education ecosystem require systematic investigation. This study employs bibliometric methods to analyze 242 publications from 2015--2025, constructing knowledge maps to reveal the evolutionary trajectory. The findings show that the field has transformed from intelligent tutoring systems to inquiry-based learning and computational thinking cultivation driven by LLMs. AI's key contribution lies in providing intelligent scaffolding that lowers the threshold for understanding knowledge. In this sense, AI is a core driving force promoting its shift from knowledge transmission to capability development.
\end{abstract}

\section{Introduction}

STEM education is undergoing a profound transformation driven by AI technology. In particular, breakthroughs in large language models (LLMs) demonstrate significant potential in optimizing instructional design, constructing personalized learning environments, and cultivating critical thinking and problem-solving skills \citep{lin2025}. However, challenges remain in integrating technology with curriculum, restructuring pedagogical models, and cultivating emerging AI literacy among both teachers and students \citep{yang2025}.

Existing reviews have explored chatbots in scientific computing education \citep{groothuijsen2024}, AI teaching tools for secondary school students \citep{crompton2024}, and AI for interdisciplinary learning \citep{cai2025}. These studies often focus on qualitative analysis or a single technological application, lacking a macroscopic knowledge map of AI across STEM education. Therefore, this study adopts a bibliometric approach to analyze core literature from the Scopus database between 2015 and 2025, aiming to answer the following research questions:

\begin{enumerate}[label=\textbf{RQ\arabic*:},leftmargin=*]
  \item What are the current core research hotspots and directions for AI in the field of STEM education?
  \item How has research output on AI in STEM education evolved over the past decade?
\end{enumerate}

\section{Methodology and materials}

\subsection{Research methodology}

As a scientific research method, bibliometrics utilizes metrical concepts and tools to analyze literature data \citep{jing2024}. This method can objectively track the developmental trajectory of a discipline, predict future research directions, and provide important references for academic planning \citep{wang2025a}. This study employs the VOSviewer software for data analysis, conducting keyword co-occurrence analysis and research theme evolution analysis on literature in the field of AI in STEM education.

\subsection{Data source and filtering}

The literature search was conducted in the Scopus database. The search query was:

\begin{quote}
\small\ttfamily\raggedright
TITLE-ABS-KEY(steam OR stem) AND TITLE-ABS-KEY(AI OR AI-Agent OR GPT OR LLM) AND TITLE-ABS-KEY(educat OR learn OR teach OR instruct).
\end{quote}

\begin{figure}[htbp]
  \centering
  \includegraphics[width=\textwidth]{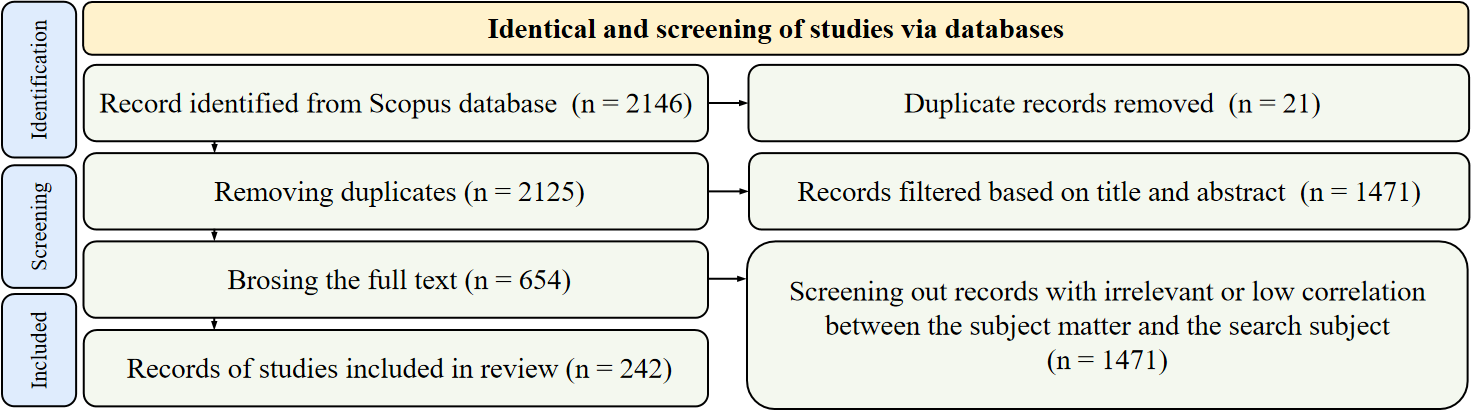}
  \caption{PRISMA flowchart for including studies to review.}
  \label{fig:prisma}
\end{figure}

The time span was set from 2015 to 2025, limited to English-language publications, yielding 2,146 documents (2,125 after deduplication). The literature was screened by excluding non-research literature and studies with low relevance to the research topic. Following this screening process (Figure~\ref{fig:prisma}), 242 core documents were identified as the analytical sample.

\FloatBarrier

\section{Bibliometric results and analysis}

\subsection{Trends in publications and disciplinary distribution}

\begin{figure}[htbp]
  \centering
  \includegraphics[width=0.94\textwidth]{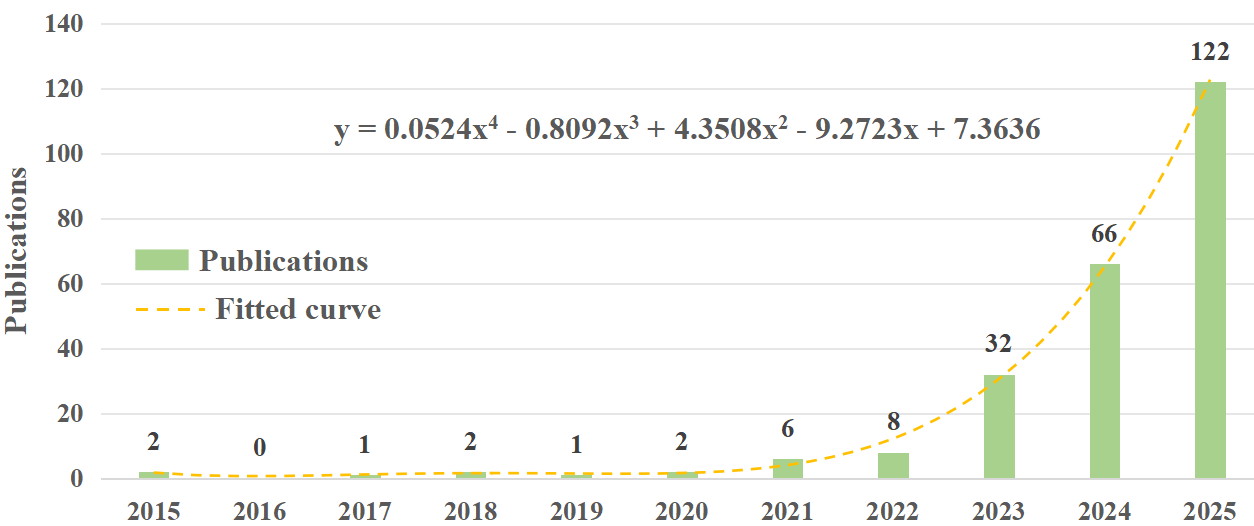}
  \caption{Trend chart of publication volume.}
  \label{fig:publication-trend}
\end{figure}

The annual publications on AI in STEM education show an upward trend (Figure~\ref{fig:publication-trend}). From 2015 to 2018, growth was moderate, representing a period of technological exploration. Since 2019, as key AI technologies matured, publication volume entered a rapid upward trajectory. Particularly after 2022, driven by the widespread implementation of LLMs, annual output showed explosive growth, maintaining strong momentum in 2024. The fitted curve suggests that research in this field is expected to maintain continued growth.

\subsection{Research hotspots and evolutionary trends}

\subsubsection{Research hotspot analysis}

\begin{figure}[htbp]
  \centering
  \includegraphics[width=0.96\textwidth]{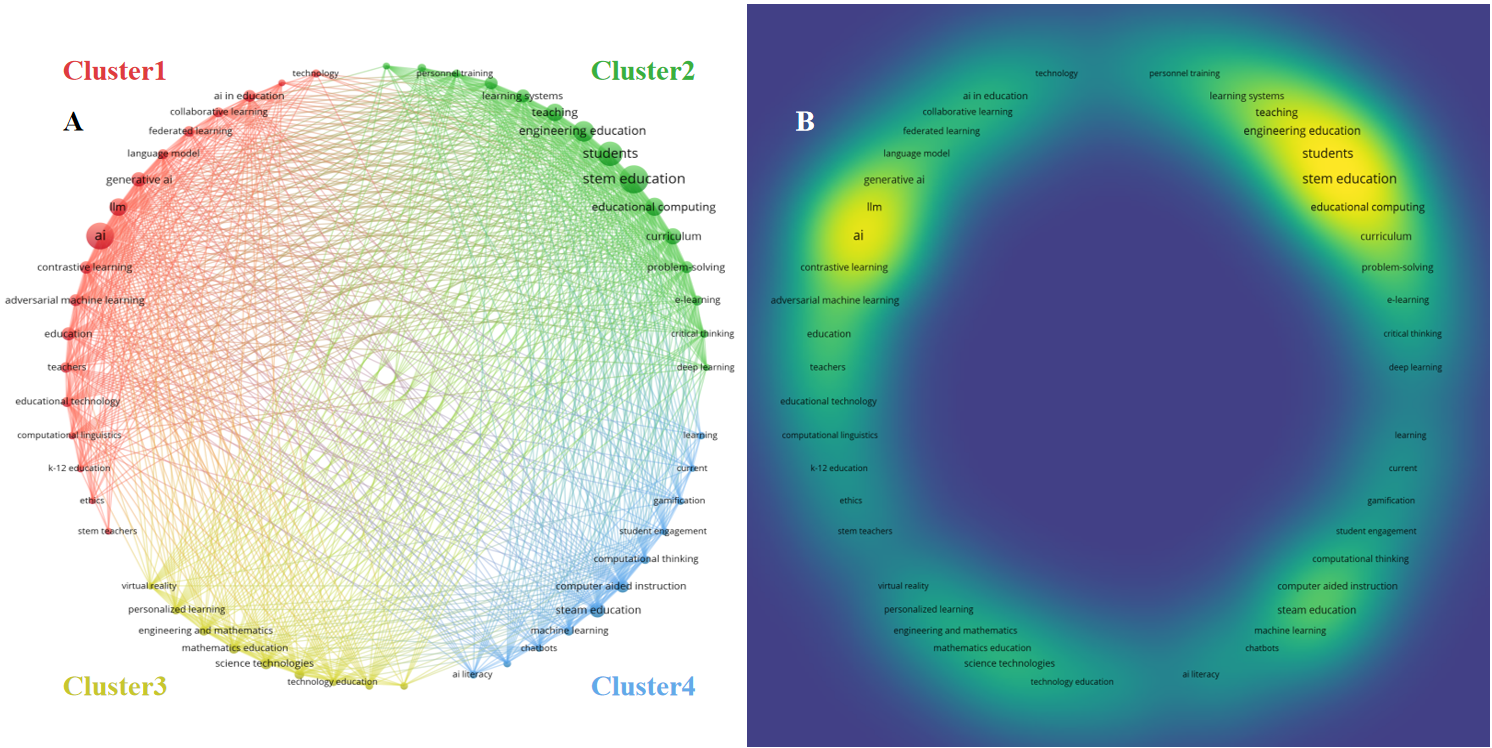}
  \caption{Keyword clustering mapping. (a) Co-occurrence clustering network. (b) Keyword density.}
  \label{fig:keyword-clusters}
\end{figure}

To explore the research hotspots of AI in the field of education, this study employed VOSviewer software to conduct a keyword co-occurrence analysis. The study identified a keyword knowledge map for research in the AI in education field. By combining the keyword co-occurrence clustering network (Figure~\ref{fig:keyword-clusters}a) and the keyword density (Figure~\ref{fig:keyword-clusters}b), we conducted analysis and discussion of the following four key cluster themes.

\FloatBarrier

\subsubsection*{Cluster 1 -- Foundational Empowerment of STEM Education by AI Breakthroughs}

This cluster examines how LLMs restructure the foundational framework of STEM education. AI facilitates the elucidation of complex scientific concepts and enhances learning outcomes. As \citet{baillifard2025} demonstrated, AI can serve as an intelligent tutor, interacting with students to improve academic performance. The incorporation of distributed techniques such as federated learning further advances sustainably evolving educational models. The FedMap model proposed by \citet{zhang2024} enables collaborative construction of learning outcomes while preserving student data privacy. These technologies underpin STEM education's transformation toward personalization and adaptivity. The keyword ``ethics'' within this cluster also reflects scholarly attention to the ethical governance and boundaries of AI application.

\subsubsection*{Cluster 2 -- Constructing STEM Pedagogical Models for Higher-Order Thinking}

As global educational goals shift toward competency development, this cluster reveals two pathways bridging theory and practice. At the curriculum design level, research advocates integrating engineering education's problem-solving orientation into curricula. \citet{guo2024} designed a curriculum for scientific data literacy in STEM disciplines, translating competency objectives into actionable instructional blueprints. At the implementation level, active learning methods enable students to actively develop their analytical, evaluative, and creative capacities \citep{halawa2024}. This signals that STEM education is progressively establishing a learner-centered model for cultivating higher-order competencies.

\subsubsection*{Cluster 3 -- Intelligent Technologies Expanding STEM Education Boundaries}

This cluster examines how intelligent technologies expand the application scenarios of STEM education along two pathways. First, immersive technologies extend instruction from physical spaces to virtual environments. Virtual reality enables learners to observe abstract scientific phenomena in an embodied manner, overcoming cognitive barriers \citep{jiang2025}. Second, adaptive technologies centered on personalized learning dynamically adjust instructional content by analyzing students' cognitive profiles and learning progress, providing precise support for individual learning pathways \citep{adewale2024}. This cluster reflects a dual exploration of ``environmental reconstruction'' and ``process restructuring,'' breaking through the spatial and organizational boundaries of STEM education.

\subsubsection*{Cluster 4 -- Technology Integration for Learning Experience\\ and Engagement Enhancement}

This cluster reveals how multi-technology integration transforms learning environments into interactive spaces. The research demonstrates that machine learning serves as the analytical core, providing decision support for personalized learning, while chatbots and gamification function as front-end interactive vehicles---particularly through multi-agent systems that construct dynamic socialized learning scenarios (Wang et al., 2025). Through human-AI interaction, students naturally develop metacognitive inquiries about ``how AI works,'' thereby transforming AI literacy cultivation into an intrinsic element of the learning experience.

\subsubsection{Research on the evolutionary trajectory}

The time-zone evolution map (Figure~\ref{fig:timezone-map}) reveals that research themes in the field of AI in STEM education experienced three evolutionary stages between 2015 and 2025. The early phase (2015--2018) focused on foundational concepts (``learning systems'' and ``machine learning''), remaining at the proof-of-concept level. The middle phase (2018--2022) saw the emergence of technical terms (``augmented reality'' and ``chatbots'') alongside contextual terms (``curriculum'' and ``engineering education,''), marking a shift from technical validation to classroom application. The recent phase (2022--2025) centers on ``generative AI'' with frequent appearances of ``personalized learning'' and ``student engagement,'' reflecting a trend toward AI-driven adaptive instruction. Based on this trajectory, the study identifies three pathways of AI-empowered STEM education:

\begin{figure}[htbp]
  \centering
  \includegraphics[width=0.98\textwidth]{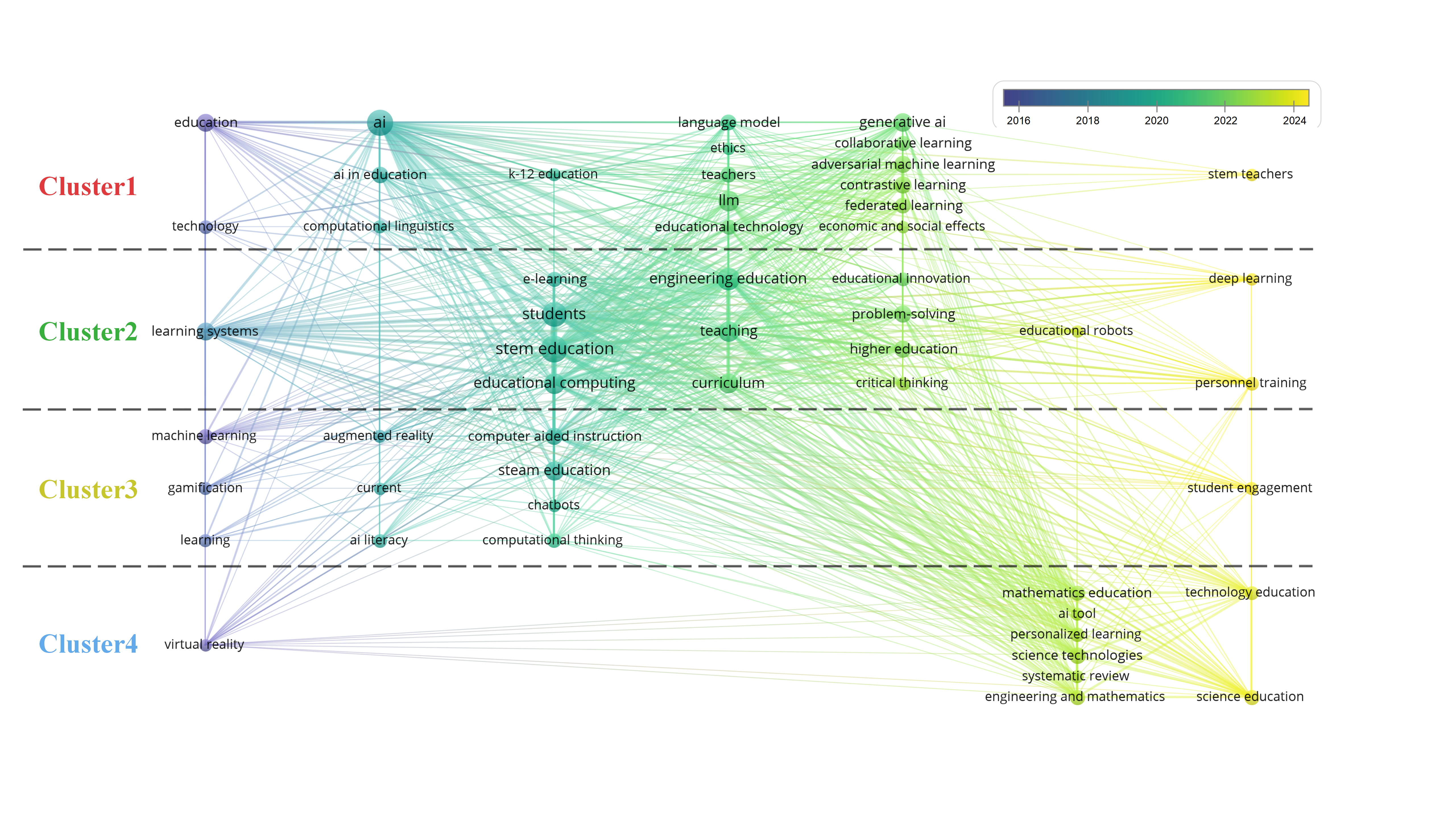}
  \caption{Time-zone evolution map of keywords.}
  \label{fig:timezone-map}
\end{figure}
\FloatBarrier

\subsubsection*{Path 1 -- The technological leap from basic AI to Generative AI}

As a core node, ``AI'' was initially linked to foundational concepts such as ``education'' and ``technology,'' then evolved toward ``generative AI,'' reflecting its transition from a basic tool to a deeply integrated instrument. Technically, breakthroughs in ``computational linguistics'' and ``LLM'' laid the foundation for GAI's educational applications, as GenAI can deepen teachers' content understanding and facilitate interdisciplinary integration \citep{zhu2025}. From the demand side, the emphasis on ``critical thinking'' and ``problem-solving'' has driven the development of novel intelligent tools, such as MindScratch, a multimodal GAI-powered visual programming tool supporting creative learning and computational thinking \citep{chen2025}. AI is evolving from a support tool to an intelligent partner deeply embedded in instruction, advancing STEM education toward personalization.

\subsubsection*{Path 2 -- Evolution from single-point technology application to integrated instructional systems}

Early-stage technologies such as ``augmented reality,'' ``gamification,'' and ``virtual reality'' formed relatively independent domains. These gradually connected through bridging concepts like ``chatbots'' and ``computer aided instruction,'' achieving deep convergence at nodes such as ``educational robots'' and ``deep learning.'' This reflects a shift from tool-oriented to system-oriented research: early studies validated individual technologies independently \citep{ates2025}, but researchers increasingly recognized that isolated innovation cannot drive substantive educational change. The focus thus shifted toward synergistic mechanisms, with AR now working alongside AI algorithms, adaptive systems, and educational robots to form intelligent instructional environments, pointing toward integrated pedagogical ecosystems.

\subsubsection*{Path 3 -- The value shift from `how to teach' to core competency cultivation}

Early research linked ``students'' and ``teachers'' primarily to ``learning systems,'' centering on knowledge transmission, with educational agents serving as recipients of technology. As the field matured, dense associations emerged with competency-oriented concepts such as ``critical thinking,'' ``problem-solving,'' and ``computational thinking,'' signaling a shift from ``how to teach'' to ``what competencies to cultivate.'' The focus turned toward higher-order competencies: inquiry-based STEM activities significantly enhance students' problem-solving skills and STEM awareness \citep{karamustafaoglu2023}. This value shift points toward personalized competency development paths beyond standardized knowledge mastery.

\section{Discussion and outlook}

\begin{figure}[htbp]
  \centering
  \includegraphics[width=0.94\textwidth]{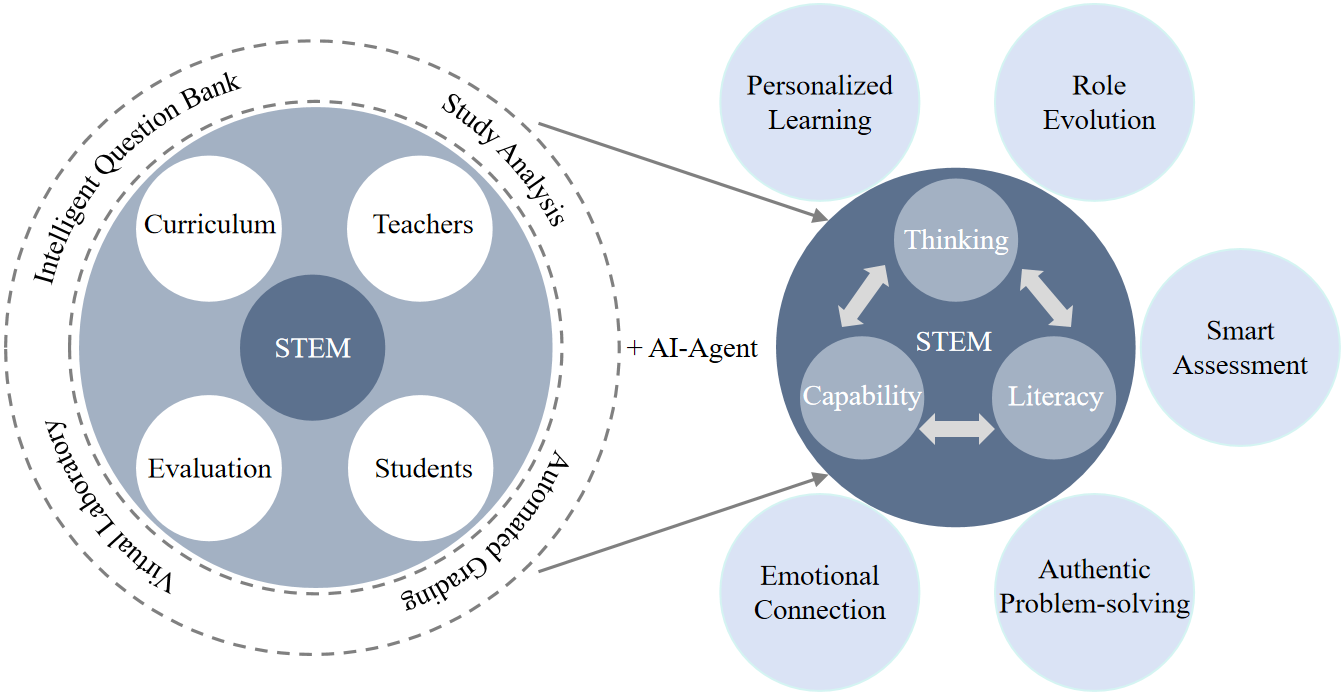}
  \caption{Conceptual framework for the AI-driven paradigm shift in STEM education.}
  \label{fig:conceptual-framework}
\end{figure}

To systematically illustrate how AI drives the paradigm shift in STEM education, this study constructs a conceptual framework (Figure~\ref{fig:conceptual-framework}) based on the preceding literature analysis. This framework is used to contrast the traditional knowledge-transmission ecosystem in STEM education with the emerging AI-driven one.

\FloatBarrier

In conventional STEM education, ``Curriculum,'' ``Teachers,'' ``Students,'' and ``Evaluation'' constitute the core elements with relatively fixed relationships. Technology is positioned as a peripheral tool, reinforcing existing teaching processes rather than fundamentally changing the knowledge-centered systemic logic.

Driven by AI, the core objectives shift into three interrelated dimensions: ``Thinking,'' ``Capability,'' and ``Literacy,'' moving the focus from ``how to teach and learn'' to systematically cultivating critical thinking, problem-solving skills, and AI literacy. AI facilitates this transition by situating learning in complex, real-world contexts and supporting adaptive learning paths to match each student's individual pace and style \citep{zhang2025}, while promoting a shift in assessment toward being more dynamic. These interconnected changes work in concert to support the differentiated development of learners, pointing toward a future in which STEM education moves beyond standardized knowledge mastery to cultivate personalized competency pathways.

\section{Conclusion}

This study analyzes the knowledge evolution and conceptual framework of AI in STEM education. The research finds that AI is profoundly reconstructing the STEM education ecosystem: the four traditional elements of curriculum, teachers, students, and assessment have undergone digital transformation through technologies such as intelligent question banks and virtual laboratories. The introduction of AI has further shifted the educational focus from knowledge transmission to the cultivation of thinking, capabilities, and competencies, while expanding the boundaries of STEM education. However, due to data sources being limited to the Scopus database, future research could further expand data sources to obtain a more comprehensive perspective.


\clearpage
\begin{thebibliography}{99}
\small
\raggedright

\bibitem[Adewale et al.(2024)]{adewale2024}
Adewale, O. S., Agbonifo, O. C., Ibam, E. O., Makinde, A. I., Boyinbode, O. K., Ojokoh, B. A., \ldots{} \& Olatunji, S. O. (2024). Design of a personalised adaptive ubiquitous learning system. \emph{Interactive Learning Environments}, \emph{32}(1), 208--228.

\bibitem[Ate\c{s} and Polat(2025)]{ates2025}
Ate\c{s}, H., \& Polat, M. (2025). Leveraging augmented reality and gamification for enhanced self-regulation in science education. \emph{Education and Information Technologies}, 1--32.

\bibitem[Baillifard et al.(2025)]{baillifard2025}
Baillifard, A., Gabella, M., Lavenex, P. B., \& Martarelli, C. S. (2025). Effective learning with a personal AI tutor: A case study. \emph{Education and Information Technologies}, \emph{30}(1), 297--312.

\bibitem[Cai et al.(2025)]{cai2025}
Cai, C., Zhu, G., \& Ma, M. (2025). A systemic review of AI for interdisciplinary learning: Application contexts, roles, and influences. \emph{Education and Information Technologies}, \emph{30}(7), 9641--9687.

\bibitem[Chen et al.(2025)]{chen2025}
Chen, Y., Xiao, S., Song, Y., Li, Z., Sun, L., \& Chen, L. (2025). MindScratch: A Visual Programming Support Tool for Classroom Learning Based on Multimodal Generative AI. \emph{International Journal of Human--Computer Interaction}, 1--19.

\bibitem[Crompton et al.(2024)]{crompton2024}
Crompton, H., Jones, M. V., \& Burke, D. (2024). Affordances and challenges of artificial intelligence in K--12 education: A systematic review. \emph{Journal of research on technology in education}, \emph{56}(3), 248--268.

\bibitem[Groothuijsen et al.(2024)]{groothuijsen2024}
Groothuijsen, S., van den Beemt, A., Remmers, J. C., \& van Meeuwen, L. W. (2024). AI chatbots in programming education: Students' use in a scientific computing course and consequences for learning. \emph{Computers and Education: Artificial Intelligence}, \emph{7}, 100290.

\bibitem[Guo et al.(2024)]{guo2024}
Guo, Q., Chen, Y., Qiao, C., \& Yu, Y. (2024). Using the WorldWide Telescope to Develop Science Data Literacy in STEM Undergraduates: A Conceptual Framework and Course Design. \emph{Journal of Science Education and Technology}, \emph{33}(6), 954--969.

\bibitem[Halawa et al.(2024)]{halawa2024}
Halawa, S., Lin, T. C., \& Hsu, Y. S. (2024). Exploring instructional design in K--12 STEM education: a systematic literature review. \emph{International Journal of STEM Education}, \emph{11}(1), 43.

\bibitem[Jiang et al.(2025)]{jiang2025}
Jiang, H., Zhu, D., Chugh, R., Turnbull, D., \& Jin, W. (2025). Virtual reality and augmented reality-supported K--12 STEM learning: trends, advantages and challenges. \emph{Education and Information Technologies}, 1--37.

\bibitem[Jing et al.(2024)]{jing2024}
Jing, Y., Wang, C., Chen, Y., Wang, H., Yu, T., \& Shadiev, R. (2024). Bibliometric mapping techniques in educational technology research: A systematic literature review. \emph{Education and Information Technologies}, \emph{29}(8), 9283--9311.

\bibitem[Karamustafao\u{g}lu and Pekta\c{s}(2023)]{karamustafaoglu2023}
Karamustafao\u{g}lu, O., \& Pekta\c{s}, H. M. (2023). Developing students' creative problem solving skills with inquiry-based STEM activity in an out-of-school learning environment. \emph{Education and Information Technologies}, \emph{28}(6), 7651--7669.

\bibitem[Lin et al.(2025)]{lin2025}
Lin, C. J., Lee, H. Y., Wang, W. S., Huang, Y. M., \& Wu, T. T. (2025). Enhancing reflective thinking in STEM education through experiential learning: The role of generative AI as a learning aid. \emph{Education and Information Technologies}, \emph{30}(5), 6315--6337.

\bibitem[Wang et al.(2025c)]{wang2025c}
Wang, H., Wang, C., Chen, Z., Liu, F., Bao, C., \& Xu, X. (2025c). Impact of AI-agent-supported collaborative learning on the learning outcomes of University programming courses. \emph{Education and Information Technologies}, 1--33. \url{https://doi.org/10.1007/s10639-025-13487-8}

\bibitem[Wang et al.(2025a)]{wang2025a}
Wang, H., Xiao, Z., Lai, Y., Bao, C., \& Wang, C. (2025a). Knowledge Mapping and Trend Analysis of AI Agents in Education: A Bibliometric Study from 2004--2024. In \emph{Proceedings of the 19th International Conference of the Learning Sciences--ICLS 2025} (pp.~1404--1408). International Society of the Learning Sciences. \url{https://doi.org/10.22318/icls2025.764364}

\bibitem[Wang et al.(2025b)]{wang2025b}
Wang, Y., Wang, Y., Tian, F., Ma, J., \& Jin, Q. (2025b). Intelligent games meeting with multi-agent deep reinforcement learning: a comprehensive review. \emph{Artificial Intelligence Review}, \emph{58}(6), 165.

\bibitem[Yang et al.(2025)]{yang2025}
Yang, Y., Sun, W., Sun, D., \& Salas-Pilco, S. Z. (2025). Navigating the AI-Enhanced STEM education landscape: a decade of insights, trends, and opportunities. \emph{Research in Science \& Technological Education}, \emph{43}(3), 693--717.

\bibitem[Zhang et al.(2024)]{zhang2024}
Zhang, Y., Li, Y., Wang, Y., Wei, S., Xu, Y., \& Shang, X. (2024). Federated learning-outcome prediction with multi-layer privacy protection. \emph{Frontiers of Computer Science}, \emph{18}(6), 186604.

\bibitem[Zhang et al.(2025)]{zhang2025}
Zhang, Y., Wang, H., \& Bao, C. (2025). Breaking the boundaries of conventional vocabulary learning: the impact of vocabulary learning with intelligent interactive companion in CAVL environments on learning outcomes, motivation, and flow experience. \emph{Computer Assisted Language Learning}, 1--28. \url{https://doi.org/10.1080/09588221.2025.2573003}

\bibitem[Zhu et al.(2025)]{zhu2025}
Zhu, Y., Jiang, H., \& Chugh, R. (2025). Empowering STEM teachers' professional learning through GenAI: The roles of task-technology fit, cognitive appraisal, and emotions. \emph{Teaching and Teacher Education}, \emph{167}, 105204.

\end{thebibliography}
\end{document}